# Enhanced Security of Symmetric Encryption Using Combination of Steganography with Visual Cryptography

Sherief H. Murad [#1], Amr M. Gody [#2], Tamer M. Barakat [#3]

[1, 2, 3] *Electrical Engineering Department, Faculty of Engineering, Fayoum University, Fayoum, Egypt*

**Abstract**
       *Data security and confidentiality is required when communications over untrusted networks take place. Security tools such as cryptography and steganography are applied to achieve such objectives, but both have limitations and susceptible to attacks if they were used individually. To overcome these limitations, we proposed a powerful and secured system based on the integration of cryptography and steganography. The secret message is encrypted with blowfish cipher and visual cryptography. Finally, the encrypted data is embedded into two innocent cover images for future transmission. An extended analysis was made to prove the efficiency of the proposed model by measuring Mean-Square-Error (MSE), Peak-Signal-to-noise-Ratio (PSNR), and image histogram. The robustness was examined by launching statistical and 8-bit plane visual attacks.   The proposed model provides a secure mean to transmit or store highly classified data that could be applied to the public security sector.*

## I.   INTRODUCTION

Data is the major part of any communication system. Computer networks provide good means to transfer these data. In some cases, private data have to be sent through networks which might be insecure due to vulnerable network segments such as wireless access points or because of computer crackers. Therefore, the need to information security has increased as the types of attacks on the data networks are also increased. There are many tools applied to keep private data safe from unauthorized access. Cryptography and Steganography are the most common tools used to achieve this level of protection. Cryptography is the art and science of using mathematical functions to transform data from their clear readable form to another scrambled form to unauthorized parties [1]. Only the intended users with additional secret information, called key, can read or access the data.

On the other hand, Steganography is the art and science of hiding the existence of secret messages into another innocent source of information such as multimedia files [2].  It does not encrypt the secret messages, but only hides them. Both cryptography and steganography have their advantages and limitations, using them individually will provide a single layer of security which could be easily compromised. Integrating them together will provide additional layers of security and their limitations will be reduced.

### A.  Blowfish Algorithm

       Blowfish is a symmetric 64-bit block cipher [3].    It uses a variable-key length from 32 and up to 448 bits. Blowfish is a Feistel network with 16 iterations. Two operations are forming the blowfish algorithm; key expansion and data encryption. Key expansion converts user's key into several sub-keys P-array and S-Boxes which will be used later in encryption and decryption. With total of 4168 bytes both P-array and S-boxes are described as follows [4]:
P-Array of 32 bits length for each entry:
P1, P2…, P18.
4 S-Boxes of 32 bits length for each entry:
$S_{1,0}, S_{1,1}, …, S_{1,255}$
$S_{2,0}, S_{2,1}, …, S_{2,255}$
$S_{3,0}, S_{3,1}, …, S_{3,255}$
$S_{4,0}, S_{4,1}, …, S_{4,255}$
As shown in figure 1, the Blowfish Algorithm is summarized in the following procedure:

*The input x is an 64-bit block data element*
*Divide x into two 32-bit halves: xL, xR*
*For i = 1 to 16 do*
*{*
*xL = xL $\oplus$ Pi*
*xR = F(xL) $\oplus$ xR*
*Swap xL and xR*
*}*
*Swap xL and xR (Undo the last swap.)*
*xR = xR $\oplus$ P17*
*xL = xL $\oplus$ P18*
*Recombine xL and xR*

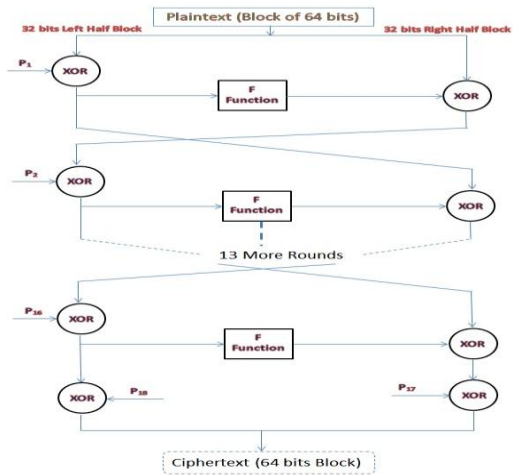

**Fig 1: Blowfish Algorithm**

Where F-Function is defined as follows:
$F(x_L) = (S_{1,a} + S_{2,b} \bmod 2^{32}) \text{ XOR } S_{3,c} + S_{4,d} \bmod 2^{32}$
Where a, b, c and d are *xL* divided into four 8-bit quarters. The additions in the above equation are modulo $2^{32}$ additions.

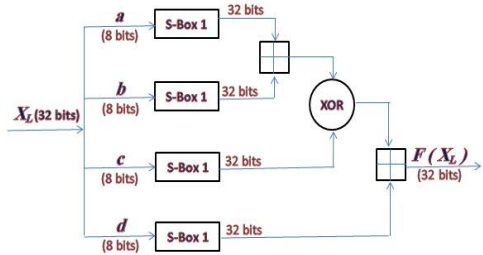

**Fig 2: F-Function used in Blowfish**

Like the other Feistel network ciphers, blowfish uses the generated sub-keys in reversed order to decrypt ciphertext.

### B. Visual Cryptography

A. Shamir and M. Naor [13] proposed a new cryptographic scheme based on secret sharing namely, visual cryptography. In (k, n) visual cryptographic scheme, a dealer encodes a secret image into *n* shares and gives each participant a share, where each share is printed on a transparency. The secret will be revealed if any *k* or more of the permitted shares stacked together. No clue about the original secret will be revealed if fewer than *k* shares are used. As illustrated in figure 3, an (2, 2) visual cryptographic scheme generates two shares. Both shares are required in order to reveal the secret back,

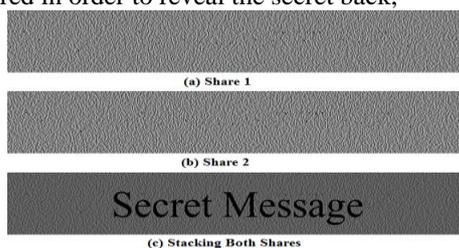

**Fig 3: Visual Cryptography**

To share a white pixel, the dealer randomly chooses one of the upper rows combinations of black/white sub-pixels, and to share a black pixel, the dealer randomly chooses one of lower rows, Figure 4. The decryption is done by stacking sub-pixels together. This process is equivalent to the OR operation to these sub-pixels. XOR operation might be used to enhance the contrast of the superimposing of the two shares. The result of stacking sub-pixels is the shown in the right column of figure 4 [14].

| Input Pixel | Sub-Pixels in Share1 | Sub-Pixels in Share2 | Superimposing of the Shares |
|---|---|---|---|
| ▢ | ▌ | + ▐ | = █ |
| ▢ | ▐ | + ▐ | = ▐ |
| ■ | ▌ | + ▐ | = █ |
| ■ | ▐ | + ▌ | = █ |

**Fig 4: Encryption process of VCS.**

### C. LSB Image Steganography

Steganography hides the secret message into another cover message. It can be applied to computer files such as document files, images, applications and communication protocols. Media files are good choice as cover objects [2]. As shown in figure 5, any steganographic system consists of three main parts, the secret message *m* to be hidden, the cover object *C* which is the carrier medium to hide *m*, and the Stego-object *S* the output of the embedding process. Least Significant Bit (LSB) algorithm conceals the secret message into cover medium through altering every LSB of the cover pixels with each bit of the secret bits [15]. As example to hide a secret message '*S*' into an image will look like this:
- *Secret message*: S
- *ASCII Value of 'S'*: 115
- *Binary representation*: 1110011
- *Binary representation of the first seven pixels of the cover image*: 101101<u>0</u>, 1100000, 1100010, 1100011, 1100101, 1100101, 1100011

After embedding the secret 'S' into the cover image, the new pixels values will be:
(101101**1**, 110000**1**, 110001**1**, 110001**0**, 110010**0**, 110010**1**, 110001**1**)

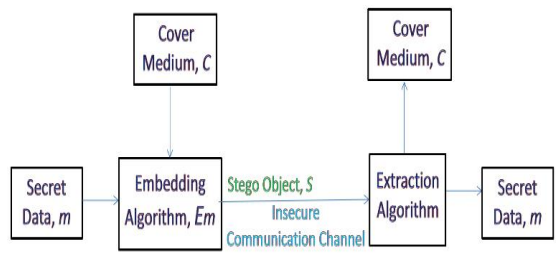

**Fig 5: LSB Steganography**

This paper is organized as follows; section 2 presents the related work. Section 3 describes the encryption and decryption phases of the proposed model. Section 4 summarizes the results and discusses the performance. Finally, section 5 concludes the paper.

## II. RELATED WORK

Many researches focus on the advantages of combining both of cryptography and steganography. This hybrid approach could be classified into two main categories; the secret message is first encrypted then embedded into carrier medium. The other approach embeds the secret message firstly into the carrier medium then encrypts them.

H. Y. Atown [3], proposed a model to secure fingerprint images by first encrypting it with transposition algorithm then hides the scrambled image into a coloured cover image using LSB. This model produced an PSNR value above 50dB.

B. Prasad, K. C. Lakshmi Narayana [4], proposed another combined scheme uses genetic steganographic algorithm to hide secret data followed by visual cryptography to add extra layer of security for the system. It works only with bitmap images and the compression depends on the document size as well as the size of the cover image. Another limitation of their model is that transmitting the noisy shares will raise the suspicion of the adversary against the intercepted images.

P. H. Dixit, K. B. Waskar, U. L. Bombale [5], introduced a multilevel network security by blend of encrypting the data with blowfish algorithm then applying LSB steganography on small embedded systems. This algorithm uses biometric features of the sender as a key for encryption. It uses the iris image for generating sub-keys for blowfish algorithm. The ciphertext is to be hidden into the iris image with LSB method before sending over network to the receiver side. The main limitation noticed is sending the iris image which was used for generating the sub-keys of blowfish algorithm. Also, with the iris image it solves the visual analysis of Stego-objects, but it cannot guarantee the statistical attacks over it.

M. Gupta, D. Chauhan [6], proposed a hybrid model which uses a (2, 2) visual cryptographic scheme along with digital watermarking to hide the generated shares. It solves the flaw of visual cryptography that could happen if an attacker alters the bits of these shares with fake bit sequences by digital watermarking.

T. S. Barhoom, S. M. Abo Mousa [7], introduced a steganographic algorithm could be applied to grey-scale and coloured images based on LSB technique. The proposed scheme aims to provide more security to the hidden information, so blowfish encryption is applied before the embedding process is done.

M. D. Munjal [8], proposed a new multilayered security scheme based on applying dual steganographic algorithms LSB and discrete wavelet transform (DWT). The secret message is embedded into two cover images. It adds more security along with good quality of the final steg-image (high PSNR, low MSE). The noticed drawback is applying statistical steganalysis on the stego-image and probably extraction of the hidden message as it is embedded as a plaintext.

R. Rathod, D. Mistry, K. Patel [9], introduced another combined approach to provide more security through first encrypting the secret message with matrix reorder or pixel shuffling to produce a scrambled version of the secret message. After that, blowfish is applied for extended encryption. Then spiral LSB method is used to hide the encrypted shuffled image.

P. K. Kavya, P. Haseena [10], proposed an efficient system for fingerprint login mechanism to authenticate users. This algorithm depends on visual cryptography and video steganography along with user input biometric information (fingerprint). Blowfish cipher is used to encrypt the fingerprint image before embedding in the video cover.

## III. PROPOSED MODEL

The hybrid of cryptography and steganography provides multilevel security to the data. The proposed system is based on Blowfish symmetric key and visual cryptography encryption ciphers along with LSB substitution steganography. The following sub-sections describe the encryption and decryption phases involved in the proposed model.

### A. Encryption Phase:

The encryption phase consists of three modules E1, E2, and E3. In E1 the secret message *m* is encrypted with the Blowfish algorithm resulting ciphertext *C*. For an extra level of security, the fingerprint of the user, KIMG, is used to generate sub-keys required in encryption rounds.

*C* is then converted into an image, *ENCIMG*, which will be supplied to the second module E2, where an (2, 2) visual cryptographic scheme is applied. After the two noisy shares, *SH1* and *SH2*, are generated, they will be concealed within two innocent cover images using LSB steganographic algorithm producing two Stego-images, *STEG1* and *STEG2*.

Later these Stego-images will be sent through two different insecure communication channels to the receiver side.

As illustrated in figure 6, the encryption procedure is summarized in the following steps:

**Inputs:** (*m, k, KIMG, C1, C2*)

**Outputs:** *(STEG1, STEG2)*

**Algorithm:**
   *[STEG1, STEG2] = EBVCS (m, k, C1, KIMG, C2)*

**Step1:** Enter Secret Message, *m*
**Step2:** Read Cover images, *C1 & C2*
**Step3:** Encrypt *m* using BlowFish Algorithm
       $C=BF_k(m)$
**Step4:** Convert *C* → *ENCIMG*
**Step5:** Encrypt Using (2, 2) Visual Cryptography
       $VC_{(2,2)}(Encimg) = \begin{cases} SH1 \\ SH2 \end{cases}$
**Step6:** Embed both SH1, SH2 into C1, C2
       *STEG1= LSB (C1, sh1)*
       *STEG2= LSB (C2, sh2)*

**Step7:** Send STEG1 & STEG2 to Receiver end.

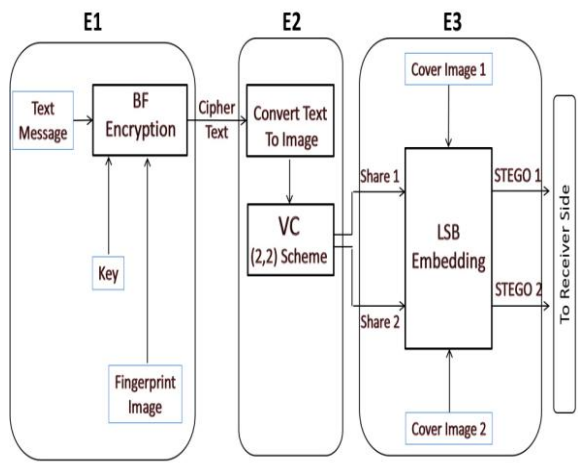

**Fig 6: The encryption phase**

### B. Decryption Phase:

The decryption phase is also consisting of three modules D1, D2, and D3. After receiving the two Stego-images, *STEG1* and *STEG2*, the D1 module extracts and reconstructs the embedded noisy shares producing *RSH1* and *RSH2*. After then, they will be provided to D2 module in which the decryption of VCS takes place to restore the ciphertext *C*.

The decryption is achieved through superimposing both shares together using the XOR operation. The other part of D2 module is the OCR function which is required to recognize *C* out of the combined image.

Finally, in D3 module, the blowfish algorithm decrypts *C* and recovers back the secret message *m*. As illustrated in figure 7, the decryption procedure is summarized in the following steps:

**Inputs:** (*k, KIMG, STEG1, STEG2*)
**Output:** *(m) secret message*

**Algorithm:**
   *[m] = DBVCS (k, KIMG, STEG1, STEG2)*

**Step1:** Receive *STEG1 & STEG2*
**Step2:** Extract Shares, *RSH1 & RSH2*
       $RSH1 = LSB^{-1} (STEG1)$
       $RSH2 = LSB^{-1} (STEG2)$
**Step3:** Combining shares → *Combined.bmp*
       $COMBINED = (RSH1 \oplus RSH2)$
**Step4:** Recognizing Text using OCR, *CTXT*
       *C= OCR (Combined)*
**Step5:** Decrypt C with Blowfish → *m*
       $m = BF_k^{-1}(CTXT)$
**Step6:** Display Results, *m* (plaintext).

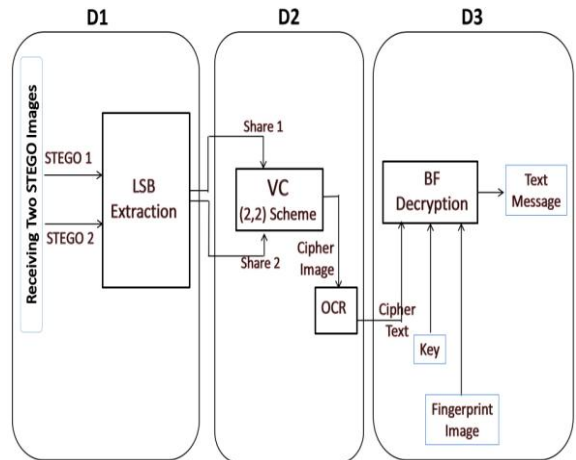

**Fig 7: The decryption phase**

### C. Application:

The main concern of the proposed model is providing more security to a highly classified data. It was implemented for public safety applications such as severs in which citizens' personal information for example, full names, addresses, and fingerprints are stored. The proposed model helps to securely transmit or store such information. The recommended configuration of the algorithm is illustrated in figure 8:

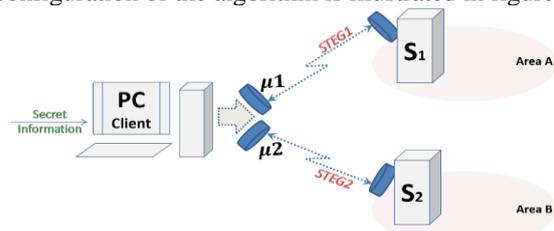

**Fig 8: The recommended Configuration**

The dealer on client machine encrypts the secret information using the encryption module of the proposed model. The information is encrypted using both blowfish then visual cryptography and finally embedded into two cover images. The output Stego-images STEG1 and STEG2 will be sent to two different storage servers S1 and S2 through point to point wireless links µ1 and µ2 at two different areas A and B respectively.

The proposed model helps in the case of stealing or breaking into servers S1 or S2 the attacker will not be able to disclose the encrypted stored images.

## IV. RESULTS AND PERFORMANCE ANALYSIS

The proposed model was implemented in MATLAB R2016b framework. As an input to the encryption phase, a text message *m,* "secret", was entered. Table-1 summarizes the results corresponding to each module E1, E2 and E3 respectively:

| |
|---|
| **(a) Secret Message:** |
| Secret |
| **(b) The first share:** |
| 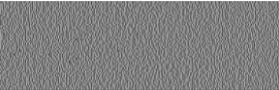 |
| **(c) The second share:** |
| 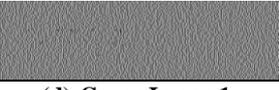 |
| **(d) Cover Image 1:** |
| 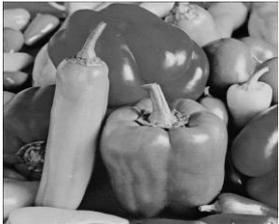 |
| **(e) Cover Image 2:** |
| 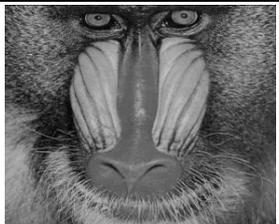 |
| **(f) STEGO Image 1:** |
| 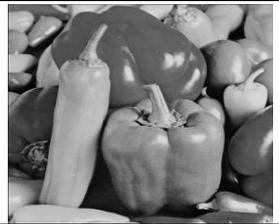 |
| **(g) STEGO Image 2:** |
| 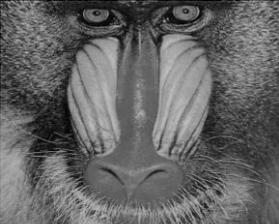 |
| **(h) Extracted RSH1:** |
| 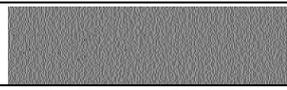 |
| **(i) Extracted RSH2:** |
| 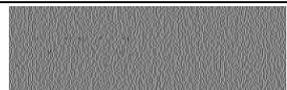 |
| **(j) Dectrypted Text:** |
| Secret |

**Table 1: The Experimental Results of the proposed model**

Quality and performance of the proposed model were measured using MSE and PSNR parameters. The results show good PSNR and MSE values. Formulas 2, 3 are used to calculate MSE and PSNR values between two images *X* and *Y*:

MSE = $\|X-Y\|^2$ / (2*N)   (2)

PSNR = 20log $((2^B-1) / \sqrt{MSE})$   (3)

For the proposed model, the PSNR range starts from 61.94 dB corresponding to a message size of 11958 bytes and reducing as the message size increased. Also, the value of the MSE starts with 0.041584 and increased as the size of the messages increased as shown in figures 9 (a) and (b).

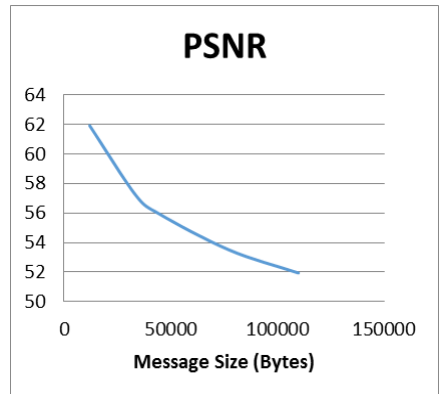

**(a)** PSNR values using different secret messages.

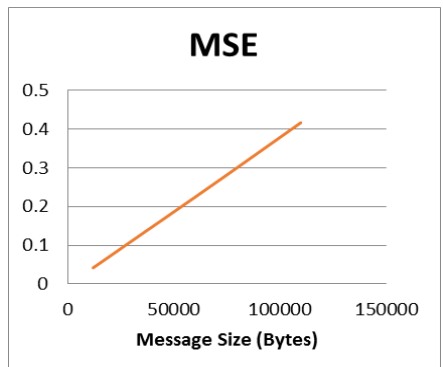

**(b)** MSE values using different secret messages.

**Fig 9: PSNR and MSE of the proposed model**

Image histogram was also used to measure to which degree the Stego-image differs from the original image. The results have shown that there is very little difference that could be noticed, as illustrated in figures 10 (a) and (b).

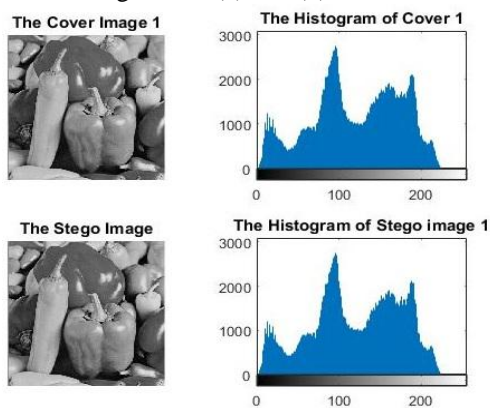

**(a) Image histograms of the Peppers.bmp.**

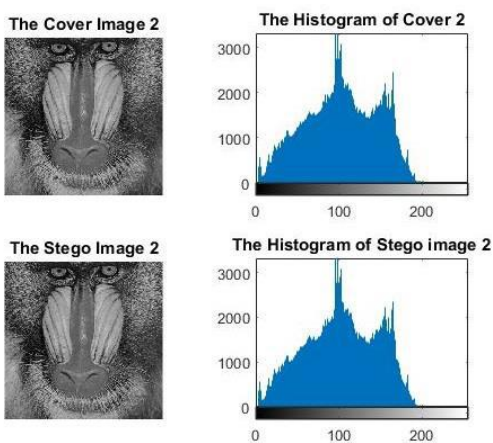

**(b) Image histograms of the Baboon.bmp.**

**Fig 10: Image Histogram of the Proposed Model**

The robustness of the proposed model was examined using statistical cryptanalysis and visual attacks. Statistical cryptanalysis was mounted using StegExpose tool [16] which is used to detect LSB embedded images. The visual attack was mounted through 8-Bitplane analysis of both cover and Stego-images [17]. The results have shown that the proposed model is resistant to these attacks.

## V. CONCLUSION

Both cryptography and steganography have limitations and could be compromised if they were applied individually. By combining them together, a strong algorithm is produced. We proposed an algorithm based on the hybrid approach through double encrypting secret data with blowfish and (2, 2) visual cryptographic then ends with LSB embedding algorithms. The produced Stego-images have perfect quality in terms of MSE and PSNR values and image histogram. The results also have shown that, the proposed model is robust to both statistical and visual steganalysis attacks. The proposed model could be applied in the public security sector to transmit or store private data.

## REFERENCES

[1] S. William, "Cryptography and network security principles and practice", 6th edition, *Prentice Hall*, New York, March 2013.
[2] Doshi, P. Jain, L. Gupta, "Steganography and Its Applications in Security", International Journal of Modern Engineering Research (IJMER), Vol.2, Pages 4634 – 4638, 2012.
[3] B. Schneier, "Description of a new variable-length key, 64-bit block cipher (blowfish) fast software encryption", Cambridge security workshop proceedings (December 1993), springer-verlag, 1994, pp. 191-204.
[4] S. Manku, K. Vasanth, "Blowfish Encryption Algorithm for Information Security", ARPN Journal of Engineering and Applied Science, Vol.10, Pages 4717-4719, 2015.M. Naor and A. Shamir, "Visual cryptography," Advances in Cryptograhy: EUROCRYPT'94, LNCS, vol. 950, pp. 1–12, 1995.
[5] S.Chandramath., Ramesh Kumar R., Suresh R. and Harish S.," An overview of visual cryptography", International Journal of Computational Intelligence Techniques, Vol.1, Pages 32-37, 2010
[6] T. Morkel, J. H. P. Eloff, M. S. Olivier, "An Overview of Image Steganography", in Proceedings of the Fifth Annual Information Security South Africa Conference (ISSA2005), Sandton, South Africa, July 2005.
[7] B. Boehm, "StegExpose A Tool for Detecting LSB Steganography", master's thesis, University of Kent, School of Computing, 2014.
[8] H. Y. Atown, "Hide and Encryption Fingerprint Image by using LSB and Transposition pixel by Spiral Method", International Journal of Computer Science and Mobile Computing (IJCSMC), Vol.3, Pages 624 – 632, 2014.
[9] S. Sekra, S. Balpande, K. Mulani, "LSB Based Steganography Using Genetic Algorithm and Visual Cryptography for Secured data Hiding and Transmission over Networks", International Journal of Computer Science and Engineering (SSRG-IJCSE), Vol.2, Pages 5 – 9, 2015.
[10] P. H. Dixit, K. B. Waskar, U. L. Bombale, "Multilevel Network Security Combining Cryptography and Steganography on ARM Platform", Journal of Embedded Systems, Vol.3, Pages 11 – 15, 2015.
[11] M. Gupta, D. Chauhan, "A Visual Cryptographic Scheme to Secure Image Shares Using Digital Watermarking", International Journal of Computer Science and Information Technologies (IJCSIT), Vol.6, Pages 4339 – 4343, 2015.
[12] T. S. Barhoom, S. M. A. Mousa, "A Steganography LSB Technique for Hiding Image Within Image using Blowfish Encryption Algorithm", International Journal of Research in Engineering and science (IJRES), Vol.3, Pages 61 – 66, 2015.
[13] M. D. Munjal, "Dual Steganography Technique Using Status LSB and DWT Algorithms", International Journal of Innovative Research in Computer and Communication Engineering (IJIRCCE), Vol.4, Pages 10483 – 10492, 2016.
[14] R. Rathod, D. Mistry, K. Patel, "Encrypted Steganography: A Combined Approach for Enhancing Image Security", International Journal of Innovative Research in Computer and Communication Engineering (IJIRCCE), Vol.4, Pages 9088 – 9096, 2016.
[15] P. K. Kavya, P. Haseena, "An Authentication Method Using A Verifible Visual Cryptography Scheme And A Steganographic Video Object Authentication Via Biometrics", International Journal of Innovative Research in Computer and Communication Engineering (IJIRCCE), Vol.6, Pages 470 – 477, 2017.
[16] B. Boehm, "StegExpose A Tool for Detecting LSB Steganography", master's thesis, University of Kent, School of Computing, 2014.
[17] P. Bateman, H. G. Schaathun, "Image steganography and steganalysis", Master's thesis, Department Of Computing, Faculty of Engineering and Physical Sciences,University of Surrey, Guildford, Surrey, United Kingdom, 2008